\documentclass[aps,prl,twocolumn,superscriptaddress,10pt]{revtex4}
\usepackage{epsfig}
\usepackage{amsmath}
\usepackage{amssymb}
\usepackage{amsfonts}

\def\unit#1{\mathrm{#1}}
\def\munu{{\mu\nu}}
\def\rL{{r_\Lambda}}       
\def\ri{\rho}              

\begin{document}

\title{Gravity tests in the solar system and the Pioneer anomaly}

\author{ Marc-Thierry  Jaekel}

\affiliation{Laboratoire de Physique Th\'eorique de l'ENS,
24 rue Lhomond, F75231 Paris Cedex 05 \footnote{
Centre National de la Recherche Scientifique (CNRS), Ecole Normale Sup\'{e}rieur
e (ENS), Universit\'{e} Pierre et
Marie Curie (UPMC); email:jaekel@lpt.ens.fr}}

\author{Serge Reynaud }

\affiliation{Laboratoire Kastler Brossel, case 74, Campus Jussieu,
F75252 Paris Cedex 05 \footnote{CNRS, ENS, UPMC;
email:reynaud@spectro.jussieu.fr}}

\begin{abstract}
We build up a new phenomenological framework associated with a minimal
generalization of Einsteinian gravitation theory. When linearity,
stationarity and isotropy are assumed, tests in the solar system are
characterized by two potentials which generalize respectively the Newton
potential and the parameter $\gamma $ of parametrized post-Newtonian
formalism. The new framework has the capability to account for the
Pioneer anomaly while preserving the compatibility of other gravity tests
with general relativity.



\end{abstract}

\maketitle

A number of tests of gravity have now been performed in the solar system and
they put severe constraints on deviations from general relativity.\cite{Will} 
However astrophysical and cosmological observations show anomalies,
notably in the rotation curves of galaxies and in the relation between
redshifts and luminosities for supernovae. Since gravity tests agree with
general relativity, these anomalies are commonly accounted for by
introducing dark components to the content of the Universe. As long as these
dark components are not detected by other means, the anomalies can also be
ascribed to modifications of standard gravity at galactic or cosmic scales.%
\cite{ModGrav} Obviously, any modification of this kind also has to match
the gravity tests in the solar system. The Pioneer anomaly may be a central
piece of information in this debate by pointing at some anomalous behaviour
of gravity at scales of the order of the size of the solar system.

The anomaly is recorded on radio tracking data from the Pioneer 10/11 probes
during their travel to the borders of the solar system.\cite{Anderson98}
The Doppler residuals, that is the differences between observed Doppler 
velocities and modelled ones, vary linearly with time, for heliocentric 
distances ranging from 20 to 70 astronomical units ($\unit{AU}$). Equivalently,
the anomaly can be described as a roughly constant acceleration $a_{P}\simeq
8\times 10^{-10}\unit{m}\unit{s}^{-2}$ directed towards the Sun. The effect
has not been explained to date though a number of mechanisms have been
considered to this aim, ranging from systematic effects to new theoretical
approaches.\cite{Anderson02,Anderson02ff,Bertolami} The potential
importance of the Pioneer anomaly for fundamental physics and space
navigation justifies it to be submitted to further experimental and
theoretical scrutiny.

In the present letter, we will focus the attention on the key question of
compatibility of the Pioneer anomaly with other gravity tests. The anomaly
cannot be explained simply from a long-range modification of the Newton
potential.\cite{Anderson02} 
If the anomalous acceleration $a_{P}$ is ascribed to such a deviation, 
its value is indeed too large to remain unnoticed on planetary tests.%
\cite{Jaekel04} 
We show in the following that this problem is cured by considering a
natural extension of general relativity. 
This extension preserves the geometrical foundations of the latter, that 
is to say the metric character of the theory, its gauge invariance, the 
law of geodesic motion and, therefore, the principle of equivalence. 
Only the dynamical equation of the metric determined by the relation
between curvature and stress tensors is modified.
The motivations of this modification will be discussed as well as its 
potential phenomenological consequences in the outer solar system.

To this aim, we may use the assumptions of linearity, stationarity and 
isotropy. As a matter of fact, the effects associated with non linearity 
of general relativity, with rotation and non sphericity of the Sun, are 
small in the outer solar system. More precisely, we will consider that these
small effects are properly taken into account in the standard description
(see for example the general relativistic calculation of Doppler data for
Pioneer probes\cite{Anderson02}) and that they have the same linear effect 
in the extended theory. As a consequence, they should disappear when 
subtracting the standard result from the modified one, so that 
the potential anomalies can be calculated with the simplifications
of a linearized theory of gravity and a stationary and isotropic source.

The phenomenological consequences of the extension will be characterized
by two potentials accomodating the phenomena usually associated with a
long-range modification of the Newton potential\cite{Fischbach} and an
Eddington parameter $\gamma$ differing from unity in the \textquotedblleft
parametrized post-Newtonian\textquotedblright\ (PPN) formalism.\cite{Will}
We will show that this opens free space to account for
the Pioneer anomaly while preserving compatibility with planetary tests. 
The extended framework also has consequences for deflection experiments
since it leads to a range dependence of the Eddington parameter $\gamma$.
This possibility, briefly discussed in the end of the letter, will be
analyzed in greater detail elsewhere\cite{LongerPaper} and shown to open 
the way to a comparison of Pioneer observations and data of most recent
deflection experiments.\cite{Bertotti} 

As a first step, we now present the motivations for the extended framework. 
As already stated, the basic geometric features of general
relativity are left unchanged. Motions are defined as Riemann geodesics
associated with a metric tensor $g_\munu =\eta _\munu +h_\munu $
with $\eta _\munu $ the Minkowski metric (signature $+1,-1,-1,-1$) and 
$h_\munu $ a small perturbation. 
The Riemann tensor $R_{\mu \rho \nu \sigma }$, the Ricci tensor $R_\munu
=\eta ^{\rho \sigma }R_{\mu \rho \nu \sigma }$ and the scalar curvature 
$R=\eta ^\munu R_\munu $ are expressed at first order in $h_\munu $
which is conveniently written in the Fourier domain, with $k$ the wavevector. 
The Einstein tensor $E_\munu =R_\munu -{\frac{1}{2}}\eta _\munu R$ 
is transverse ($k^{\nu }E_\munu=0 $) as a consequence of Bianchi identity, 
and this is also the case for the stress tensor $T_\munu $ 
due to energy-momentum conservation.

In Einsteinian theory, these two tensors are merely proportional to each
other $E_\munu =\frac{8\pi }{c^{4}}G_{N}T_\munu $ with $G_{N}$ the
Newton constant. But it is easy to write a more general linear relation
between the two tensors 
\begin{equation}
E_\munu \left[ k\right] =R_\munu -{\frac{1}{2}}\eta _\munu
R=\chi _\munu {}^{\rho \sigma }\left[ k\right] T_{\rho \sigma }\left[ k%
\right]  \label{Linear}
\end{equation}%
$\chi _\munu {}^{\rho \sigma }$ describes a momentum-dependent linear
response of spacetime curvature to stress tensors. This relation preserves
transversality of $E_\munu $ which can be read as a constraint $k^{\nu
}\chi _\munu {}^{\rho \sigma }=0$ on the linear susceptibility. In spite
of this constraint, the susceptibility can still take different forms as
proven by introducing projections over traceless and traced components \cite%
{Jaekel} 
\begin{eqnarray}
E_\munu  &=&E_\munu ^{(0)}+E_\munu ^{(1)} \\
E_\munu ^{(0)} &=&\left( \frac{\pi _{\mu }^{\rho }\pi _{\nu }^{\sigma
}+\pi _{\mu }^{\sigma }\pi _{\nu }^{\rho }}{2}-\frac{\pi _\munu \pi
^{\rho \sigma }}{3}\right) E_{\rho \sigma }  \notag \\
E_\munu ^{(1)} &=&\frac{\pi _\munu \pi ^{\rho \sigma }}{3}E_{\rho
\sigma }\quad ,\quad \pi _\munu =\eta _\munu -\frac{k_{\mu }k_{\nu }%
}{k^{2}}  \notag
\end{eqnarray}%
The factors in front of $E_{\rho \sigma }$ in the righthand side of these
equations are orthogonal projectors on the two sectors explored by
transverse tensors. The component (0) has a null trace and it is related to
the conformally invariant Weyl curvature tensor whereas the component (1) is
related to the scalar curvature.

Linear response functions $\chi _\munu {}^{\rho \sigma }$ are naturally
produced by radiative corrections to general relativity.\cite{Jaekel} 
Such corrections may for example give rise to modifications of the 
lagrangian which behave, at lowest order, as quadratic forms of the 
curvature tensors and which differ in the two sectors of traceless and 
traced perturbations.\cite{UtiyamaDeWitt} In particular, electromagnetic 
corrections only contribute to the sector (0) as a consequence of 
conformal invariance. Effects of such corrections in the inner solar system 
have been looked for and not found.\cite{Stelle} 
Long range effects may also be expected, for example from the Sakharov argument 
deriving gravity from a kind of elasticity of quantum vacuum.\cite{Sakharov} 
Anyway, Einstein theory is not renormalizable\cite{tHooft} so that the
response functions cannot be fully calculated from the first principles. 
In the sequel of this letter, we adopt a phenomenological point of view by
considering the long range modifications allowed by equations (\ref{Linear}) 
and comparing their consequences to observations in the outer solar system.

We now go one step further by describing the Sun as a point source $T_{\rho
\sigma }\left( x\right) =\eta _{\rho 0}\eta _{\sigma 0}Mc^{2}\delta \left( 
\mathbf{x}\right) $ at rest in the center of the solar system; $M$ is the
mass of the Sun and $\delta \left( \mathbf{x}\right) $ a 3-dimensional Dirac
distribution bearing on the space coordinates $\mathbf{x}$. As a consequence
of stationarity, there is no time variation and the frequency $k_{0}$
remains null in the following. The Einstein tensor is written in Fourier
space in terms of the spatial part $\mathbf{k}$ of the wavevector $E_{\mu
\nu }\left[ \mathbf{k}\right] =\chi _\munu {}^{00}\left[ \mathbf{k}\right] 
Mc^{2}$. As a consequence of isotropy, the linear expression (\ref{Linear}) 
of $E_\munu $ takes the form 
\begin{eqnarray}
E_\munu \left[ \mathbf{k}\right] &=&\left( \frac{\pi _{\mu }^{0}\pi
_{\nu }^{0}+\pi _{\mu }^{0}\pi _{\nu }^{0}}{2}-\frac{\pi _\munu \pi ^{00}}
{3}\right) \widetilde{G}^{(0)}\left[ \mathbf{k}\right] \frac{8\pi M}{c^{2}}
\notag \\
&&+\frac{\pi _\munu \pi ^{00}}{3}\widetilde{G}^{(1)}\left[ \mathbf{k}%
\right] \frac{8\pi M}{c^{2}}  \label{postEinstein}
\end{eqnarray}%
This constitutes a twofold generalization of Einstein equation which is
recovered for $\widetilde{G}^{(0)}\left[ \mathbf{k}\right] =\widetilde{G}%
^{(1)}\left[ \mathbf{k}\right] =G_{N}$. First, the scalar functions $%
\widetilde{G}^{(0)}\left[ \mathbf{k}\right] $ and $\widetilde{G}^{(1)}\left[ 
\mathbf{k}\right] $ are momentum dependent thus having the status of running
coupling constants.\cite{Running} Then, these functions differ in the two
sectors, which will turn out to be the key point for accomodating a
Pioneer-like anomaly.

In order to write the solution of equations (\ref{postEinstein}) which is
stationary and isotropic, we use the PPN gauge\cite{Will} 
\begin{eqnarray}
&&h_{00}\left( r\right) =2\Phi _{N}\left( r\right)  \label{Potentials} \\
&&h_{jk}\left( r\right) =2\left( \Phi _{N}\left( r\right) -\Phi _{P}\left(
r\right) \right) \eta _{jk}  \notag
\end{eqnarray}%
The two potentials $\Phi _{N}$ and $\Phi _{P}$ depend on the distance $r$ to
the Sun. The first one $\Phi _{N}\left( r\right) $ represents a Newton
potential with long-range modifications.\cite{Fischbach} 
Comparison with planetary observations will constrain $\Phi _{N}$ to remain 
close to its standard expression $\phi\equiv -\frac{G_N M}{c^2r}$. 
The second potential $\Phi _{P}\left( r\right) $ promotes the PPN parameter 
$\gamma $ from the status of a constant to that of a function. 
As shown below, the presence of $\Phi _{P}$ is sensed by light, 
so that deflection experiments\cite{Bertotti} lead to constraints 
on $\Phi _{P}$ in the vicinity of the Sun. 
We will also see that the presence of $\Phi _{P}$ in the outer solar system
may affect eccentric motions, thus allowing for a possible gravitational 
interpretation of the Pioneer anomaly. 

At this point, it is worth adressing briefly the question of nonlinearity of 
gravitation theory accounted for by a parameter $\beta$ in the PPN metric.%
\cite{Will} Its effect is known to be significant in the inner part of 
the solar system by affecting the perihelion precession of Mercury 
as well as the polarization of the Moon/Earth orbit by the Sun.
Its discussion requires a non linear treatment of gravitational 
perturbations, which is clearly out of the scope of this letter.
The present discussion will thus be restricted to the category of 
phenomena correctly accounted for in the linearized treatment.
This category corresponds to probes having kinetic energies 
$v^2 /c^2$ much larger than their gravitational energies $\left|\phi\right|$,
thus including the case of Pioneer-like probes escaping the solar system 
as well as the case of light.\cite{LongerPaper}

The two potentials $\Phi _{N,P}$ are directly related to the running
gravitational constants. Writing down equations (\ref{postEinstein}) in the
PPN gauge, we indeed obtain 
\begin{equation}
\frac{\Delta \Phi _{a}\left( \mathbf{x}\right) }{4\pi }=\frac{\widetilde{G}%
_{a}\left( \mathbf{x}\right) M}{c^{2}}\quad ,\quad a=N,P
\end{equation}%
where $\Delta $ is the spatial Laplacian operator and $\widetilde{G}_{N,P}$
are determined by $\widetilde{G}^{(0,1)}$ 
\begin{equation}
\left( 
\begin{array}{c}
\widetilde{G}_{N}\left[ \mathbf{k}\right] \\ 
\widetilde{G}_{P}\left[ \mathbf{k}\right]%
\end{array}%
\right) \equiv \frac{1}{3}\left( 
\begin{array}{cc}
4 & -1 \\ 
2 & -2%
\end{array}%
\right) \left( 
\begin{array}{c}
\widetilde{G}^{(0)}\left[ \mathbf{k}\right] \\ 
\widetilde{G}^{(1)}\left[ \mathbf{k}\right]%
\end{array}%
\right)
\end{equation}%
The standard Poisson equation is recovered when $\widetilde{G}_{N}=G_{N}$ 
and $\widetilde{G}_{P}=0$ ({\it ie} $\widetilde{G}^{(0)}=\widetilde{G}^{(1)}=G_{N}$).
In the general case, the functions $\widetilde{G}_{N,P}$ are momentum dependent. 

In a simple version of the framework, the two potentials contain
contributions linear in $r$ besides the ordinary contributions scaling as 
$\frac{1}{r}$ 
\begin{equation}
\Phi _{a}\left( r\right) =-\frac{G_{a}M}{rc^{2}}+\frac{\zeta _{a}Mr}{c^{2}}
\quad ,\quad a=N,P  \label{Pot_minimal}
\end{equation}
This corresponds to the following running constants 
\begin{equation}
\widetilde{G}_{a}\left[ \mathbf{k}\right] =G_{a}
+\frac{2\zeta _{a}}{\mathbf{k}^{2}}\quad ,\quad a=N,P  \label{Minimal}
\end{equation}
$G_{N}$ is identified as the effective Newton constant in the inner solar
system while the 3 parameters $G_{P}$, $\zeta _{N}$ and $\zeta _{P}$ measure
the deviation from general relativity; $\zeta _{N}$ describes a long range
modification of Newton law which has to remain small to fit planetary data; 
$G_{P}$ corresponds to the anomaly $(\gamma-1)$ of the PPN formalism (see below);  
$\zeta _{P}$ is the newest feature of the extended framework, leading 
in the following to a Pioneer-like anomaly.

It is worth noticing that expressions (\ref{Pot_minimal}-\ref{Minimal}) cannot 
be exact for all distances or all momenta. 
What is needed for our purpose is that they are correct descriptions of the true 
running constants at the scales involved in the tests. 
In particular, the linear increase of (\ref{Pot_minimal}) with $r$ has to hold 
roughly from 20 to 70 AU in order to explain the Pioneer anomaly (see below). 
But the same increase continued at much larger scales would have unwanted 
consequences for galactic astrophysics. 
This implies that expressions (\ref{Pot_minimal}) can only be approximations
in the solar system of functions better behaved at $r\rightarrow \infty $. 
For example, they can be considered as an expansion at $r\ll\rL$ of Yukawa 
functions with a range $\rL $ larger than the size of the solar system.
Accordingly, the infrared divergence of the running constants (\ref{Minimal}) 
is cured by an infrared cutoff at a wavevector of the order of $1/\rL$. 
This regularization does not solve the problem of the matching of solar system 
physics with larger scales but it is sufficient to discuss
phenomena within the solar system. 
Anyway, (\ref{Pot_minimal}) is only a simple model for the potentials $\Phi_{N,P}$ 
with the advantage that the phenomenology is determined by a small number of 
constants but the drawback of a loss of generality with respect to the
more general $r$-dependence allowed for the metric (\ref{Potentials}).

We now discuss the potentially observable consequences of this framework.
To this aim, we study geodesic motion in the metric (\ref{Potentials}) or,
equivalently, Hamilton-Jacobi equation for wave propagation. We denote by $t$
the time coordinate, $r$ the radius, $\varphi $ the azimuth angle, 
$\theta $ the colatitude and suppose the trajectory to take place in the
plane $\theta ={\frac{\pi }{2}}$. For matter waves with a non null mass for
example, we use the conservation of energy 
$E=mc^{2}g_{00}\frac{c\mathrm{d}t}{\mathrm{d}s}$ and angular momentum 
$J=mr^{2}\mathrm{sin}^{2}\theta g_{11}\frac{\mathrm{d}\varphi }{\mathrm{d}s}$ 
where $\mathrm{d}s$ is the invariant length element along the motion. 
We also denote by 
$\upsilon _{r}\equiv c\frac{\mathrm{d}r}{\mathrm{d}s}$ and 
$\upsilon_{\varphi }\equiv cr\frac{\mathrm{d}\varphi }{\mathrm{d}s}$ 
the velocities measured respectively in the radial and orthoradial directions. 
In a first stage, we briefly discuss the effect of a modification of the first
potential $\Phi _{N}$, setting $\Phi _{P}$ to zero, and we recover the known
fact that it cannot account for the Pioneer anomaly. We then shift our
attention to the effect of $\Phi _{P}$ alone. In the concluding paragraphs,
we will sketch the program of reanalyzing gravity tests in the general
case where $\Phi _{N}$ can be modified and $\Phi _{P}$ differ from zero. 

A long range modification of the Newton potential $\Phi _{N}$  could be
detected as an anomaly of the third Kepler law on a circular orbit ($%
\upsilon _{r}=0$). To evaluate this effect, we compute the square $\upsilon
_{\varphi }^{2}$ of the orthoradial velocity in the modified theory and
subtract from it the constant value $\left[ \upsilon _{\varphi }^{2}\right]
_{\mathrm{st}}$ obtained in standard theory. We denote by $\delta \upsilon
_{\varphi }^{2}$ the difference which measures the potential anomaly  
\begin{equation}
\delta \upsilon _{\varphi }^{2}\equiv \upsilon _{\varphi }^{2}-\left[
\upsilon _{\varphi }^{2}\right] _{\mathrm{st}}{~\simeq ~}\zeta _{N}Mr
\label{Kepler}
\end{equation}%
This anomaly, which is proportional to $\zeta _{N}$, is not observed on the
motions of planets or probes in the solar system.\cite{Fischbach} In
particular, the telemetry data on probes close to Mars are sufficient to put
an upper bound on $\zeta _{N}$ which is much too small to account
for the Pioneer anomaly (see section XI-B in \cite{Anderson02}). 
After a translation into the notations of the present paper, the bound 
on the anomalous acceleration 
$\left\vert \zeta _{N}M\right\vert <5\times 10^{-13}\unit{m}\unit{s}^{-2}$ 
is found to be much smaller than $a_{P}$. 
Since a long-range modification of the Newton potential $\Phi _{N}$
cannot explain the anomaly, we now consider the case where the first
potential $\Phi _{N}$ has its standard form and focus our
attention of the effects of the second potential $\Phi _{P}$.

The potential $\Phi _{P}$ affects Doppler tracking data of Pioneer-like probes
with eccentric motions. 
To evaluate this effect, we calculate the motion of such probes in the metric 
(\ref{Potentials}) and also take into account the perturbation of the 
propagation of radio signals to and from the probes. 
We then express the result as an equivalent acceleration $a$ defined as the
time derivative of the Doppler velocity. We finally obtain the anomaly 
by subtracting the standard expression calculated with general relativity. 
The anomaly is found to be proportional to the derivative 
$\frac{\mathrm{d}\Phi _{P}}{\mathrm{d}r}$ of the second potential as well as to the 
square $\upsilon _{r}^{2}$ of the radial velocity of the probes\cite{LongerPaper} 
\begin{equation}
\delta a\equiv a-\left[ a\right] _{\mathrm{st}} 
\simeq 2\frac{\mathrm{d}\Phi _{P}}{\mathrm{d}r}\upsilon
_{r}^{2}\simeq 2\left( \zeta _{P}M+\frac{G_{P}M}{r^{2}}\right) \frac{%
\upsilon _{r}^{2}}{c^{2}}  \label{Pioneer}
\end{equation}%
We know that $G_{P}\ll G_{N}$ (see also below) and $\upsilon _{r}^{2}\ll c^{2}$, 
so that the term proportional to $G_{P}$ can be neglected in (\ref{Pioneer}). 
We are therefore left with a constant anomalous acceleration $\delta a$
directed towards the Sun, if $\zeta _{P}$ has a positive sign. 
Tentatively identifying this result with the Pioneer anomalous acceleration 
fixes the unknown parameter $\left\vert\zeta_{P}M\right\vert =%
a_{P} c^{2}/(2\upsilon _{r}^{2})\simeq 0.25\unit{m}\unit{s}^{-2}$
which is thus found to be much larger than $\left\vert \zeta _{N}M\right\vert$. 

Equation (\ref{Pioneer}) means that the new framework presented in this
letter effectively has the capability of accomodating a Pioneer-like anomaly
for probes having a large radial velocity. 
It also leads to the prediction that the anomalous acceleration should show a 
dependence versus the velocity of the probes. 
As the two Pioneer probes have nearly equal velocities and nearly equal 
accelerations, this prediction cannot be confronted to available data. 
But it could be checked out by confronting (\ref{Pioneer}) to the data 
recorded on Pioneer probes. \cite{Anderson02ff} 
This prediction also has to be kept in mind when proposing new missions, 
since it points to the idea of trying probes with different radial 
velocities. 
Equation (\ref{Pioneer}) also predicts a specific $r-$dependence for 
the anomalous acceleration when the term proportional to $G_{P}$ and 
the variation of $\upsilon _{r}^{2}$ on the trajectory are kept.
This might be tested not only in the outer solar system but also on probes 
flying to Mars or Jupiter, if the sensitivity of the acceleration
measurement can be made good enough in spite of perturbations such as solar
wind and radiation pressure.\cite{Anderson02}
This might be done by embarking an accelerometer in the probe and thus
distinguishing between drag forces and geodesic motion.

In the sequel of the letter, we discuss other phenomenological consequences 
of the presence of $\Phi _{P}$ and check out that the modification of Einstein
theory needed to obtain  (\ref{Pioneer}) does not spoil its good agreement
with other gravity tests. 
A critical problem in this context is the effect of $\Phi_{P}$ on propagation 
of light rays. In order to assess this effect, we compute the deflection
angle $\psi$ for light rays passing near the Sun in the metric (\ref%
{Potentials}) and subtract the standard value to obtain the potential
anomaly as\cite{LongerPaper} 
\begin{equation}
\delta \psi \equiv \psi -\left[ \psi \right] _{\mathrm{st}}
\simeq -\frac{2G_{P}M}{\ri c^{2}}-\frac{2\zeta _{P}M\ri}{c^{2}}
L\left( \ri \right)   \label{Eddington}
\end{equation}%
$\ri$ is the impact parameter, that is also the distance of closest approach 
of the light ray to the Sun; $L$ is a factor of order unity which depends 
logarithmically on $\ri$ and on the distances of the emitter and receiver 
to the Sun; for an emitter far outside the solar system, its distance is
replaced by the range $\rL$ at which the linear dependence of the metric 
falls down to zero. 
Should $\zeta _{P}$ be set to zero, equation (\ref{Eddington}) would be 
equivalent to the PPN result (see \cite{Will}) 
with an Eddington parameter $\gamma $ determined from the ratio of
gravity constants $G_{P}/G_{N}=( 1-\gamma )$. 
Deflection tests would thus tell us that $G_{P}/G_{N}$ is much smaller than
unity with a maximum value given by the upper bound on $(\gamma -1)$. 
The novelty in equation (\ref{Eddington}) is the term proportional to 
$\zeta _{P}$ which implies that deflection tests could show an anomaly
depending on the impact parameter.
This result certainly pleads for a reanalysis of the most recent 
deflection data obtained during the Cassini mission.\cite{Bertotti}
A comparison of these data with (\ref{Eddington}) or, preferably, the more precise
expression in \cite{LongerPaper}, would indeed lead to a test of the presence
of $\zeta _{P}$ independent of the Pioneer test.
An observation of a non null $\zeta _{P}$ would constitute a clear indication for 
the new framework. 
Note that different values of $\zeta _{P}$ could also be deduced from Pioneer and
Cassini data since $\zeta _{P}$ is in fact a function which could vary 
when going from the radius of the Sun $\sim 0.7\times 10^{9}\unit{m}$ 
to the distance explored by the Pioneer probes $\sim 1.2\times 10^{13}\unit{m}$.
On a longer term, the $\ri$-dependence of deflection experiments also constitutes
a further motivation for high accuracy deflection tests such as LATOR\cite{LATOR} 
or astrometric surveys such as GAIA.\cite{GAIA}

The present letter constitutes a first study of the phenomenological 
consequences of the extended framework. As already stated, the
modified gravity equation naturally leads to metric perturbations
characterized by two potentials $\Phi _{N}$ and $\Phi _{P}$. 
It is therefore necessary to perform a new analysis of the motions in 
the solar system looking for the combined effects of these two potentials. 
The nonlinearity of gravitation theory has also to be taken into account
in order to evaluate for example the precession of perihelion of planets.
Clearly the eccentricity of the orbits will play a key role in these
discussions; it takes large values for Pioneer-like probes which sense 
$\Phi _{P}$ whereas it is zero for circular orbits which do not.
This suggests to devote a detailed analysis to the intermediate situation,
not only for the two categories of bound and unbound orbits, but also for the 
flybies used to bring Pioneer-like probes from the former category to 
the latter one.\cite{Anderson02,Anderson02ff} 

\def\Rev#1#2#3#4{{\it #1} {\bf #2} #4 (#3)}
\def\Book#1{{\it #1}}
\def\Name#1#2{#2 #1}
\def\etal{{\it et al} }
\def\ibid{{\it ibidem} }
\def\eprint#1{{\rm #1} }

\def\PR{Phys. Rev.}
\def\PRL{Phys. Rev. Lett.}
\def\RMP{Rev. Mod. Phys.}
\def\JMP{J. Math. Phys.}
\def\MPL{Mod. Phys. Lett.}
\def\RPP{Rep. Progr. Phys.}
\def\PRep{Phys. Rep.}
\def\JPe{J. Physique}
\def\EpL{Europhys. L.}
\def\EPJ{Eur. Phys. J.}
\def\CQG{Class. Quantum Grav.}
\def\GRG{Gen. Rel. Grav.}
\def\ApJ{Astrophys. J.}
\def\AA{Astron. Astrophys.} 
\def\MNRAS{Mon. Not. R. Astr. Soc.}
\def\AJP{Am. J. Phys.}
\def\APk{Ann. Physik}
\def\SP{Sov. Phys.}
\def\IJMP{Int. J. Mod. Phys.}

\end{document}